\documentclass[11pt]{article}

\usepackage[margin=1in]{geometry}
\usepackage{amsmath,amssymb,amsthm}
\usepackage{graphicx}
\usepackage{enumitem}
\usepackage{natbib}  

\usepackage[hidelinks]{hyperref}

\newtheorem{theorem}{Theorem}[section]

\newtheorem{proposition}[theorem]{Proposition}
\newtheorem{corollary}[theorem]{Corollary}
\theoremstyle{definition}

\theoremstyle{remark}
\newtheorem{remark}[theorem]{Remark}

\setlist[enumerate]{leftmargin=.5in}
\setlist[itemize]{leftmargin=.5in} 

\title{\Large\textbf{Resonance matching of 2-$\delta$ and 3-$\delta$ potentials\\
in 1D Quantum Scattering}}

\author{Naw Sai\thanks{Department of Mathematics, Hong Kong Baptist University. Email: \texttt{nawsai1abang@gmail.com}} \\[0.5em]
}
\date{}
\begin{document}

\maketitle

\begin{abstract}
We investigate whether a 3-$\delta$ system with positive coupling strengths can approximate the transmission spectrum of a 2-$\delta$ resonance system with opposite-sign couplings for $k < 3$. Theoretical analysis establishes exact isospectrality---perfectly matched transmission spectrum---is impossible for physically non-trivial configurations, while numerical experiments identify the minimal constraint set for practicability. These results establish both the practical limits and achievable accuracy of resonance matching under sign constraints, with implications for understanding spectral non-uniqueness in quantum scattering problems. 
\end{abstract}

\section{Introduction}
Can different potentials produce identical quantum scattering spectra? This question of isospectrality lies at the heart of the inverse scattering problem, where one attempts to reconstruct a potential from its scattering data \cite{Chadan1989}. While exact isospectrality has been explored in special cases (supersymmetric partners \cite{Junker1996,Cooper1995}, Darboux transformations \cite{Matveev1991}), the practical question of \textit{approximate} isospectrality under physical constraints requires further investigation.

Delta-function potentials serve as idealized models for point scatterers in quantum mechanics \cite{Griffiths2018}, with applications in the modeling of resonant tunneling structures and quantum transmission engineering \cite{SanchezSoto2018}. We focus on the low-energy regime ($k < 3$), where resonance structures are most pronounced and potential details significantly influence transmission. At higher energies, delta-function scatterers become increasingly transparent, with transmission approaching unity asymptotically, making spectral discrimination less relevant \cite{Taylor2006}. In systems with multiple delta functions, the relative signs of the coupling strengths critically influence transmission through interference effects between scatterers. While allowing arbitrary signs enables more complex resonance structures, restricting to positive-only coupling strengths---where all delta functions have the same sign---arises naturally in certain experimental contexts and provides a well-defined mathematical constraint for exploring the limits of spectral approximation.

Given these positive-only constraints, this work investigates how 3-$\delta$ systems can approximate the transmission spectrum, particularly the resonance structure, of a 2-$\delta$ system with opposite-sign coupling strengths. Although the 3-$\delta$ system possesses an additional spatial degree of freedom, the sign constraint fundamentally restricts the solution space. The central question is whether this extra positioning flexibility can compensate for the inability to use attractive wells and at what cost to spectral fidelity. 

We approach this through complementary analytical and numerical methods. Analytically, we prove that perfect spectral matching across all energies is impossible for non-trivial configurations, establishing fundamental limits on achievable accuracy even when resonances are individually matched. Numerically, we employ windowed optimization to match resonance positions and widths by finding 3-$\delta$ configurations within localized energy ranges. Our results demonstrate that, despite this impossibility result, approximate resonance matching can be achieved with surprisingly high fidelity under minimal additional constraints on the coupling strengths.

\section{Theory}
\label{sec:framework}

\subsection{Schr\"{o}dinger Equation and Delta Potentials}

The time-independent Schr\"{o}dinger equation for a particle scattering off delta function 
potentials in 1-dimension is:
\begin{equation}
-\frac{\hbar^2}{2m}\psi''(x) + V(x)\psi(x) = E\psi(x)
\end{equation}
where the potential consists of $N$-$\delta$ functions:
\begin{equation}
V(x) = \sum_{i=1}^N \alpha_i \delta(x - x_i)
\end{equation}
Here $\alpha_i$ represents the strength of the $i$-th delta potential at position $x_i$, and 
$E = \hbar^2 k^2/2m$ ($k = \sqrt{2mE}/\hbar$) is the energy of the incident particle.

The Dirac delta function $\delta(x - x_i)$ is a generalized function defined by:
\begin{equation}
\delta(x - x_i) = \begin{cases}
+\infty & \text{if } x = x_i \\
0 & \text{if } x \neq x_i
\end{cases}
\end{equation}
with the normalization condition:
\begin{equation}
\int_{-\infty}^{\infty} \delta(x - x_i)\,dx = 1
\end{equation}
More rigorously, it is characterized by the sifting property:
\begin{equation}
\int_{-\infty}^{\infty} f(x)\delta(x - x_i)\,dx = f(x_i)
\end{equation}
for any continuous function $f(x)$.

\subsection{Transfer Matrix}

The transfer matrix method provides an efficient way to handle the matching conditions imposed by delta potentials \cite{SanchezSoto2018,Flugge1999,Merzbacher1998}. Consider a single delta potential $V(x) = \alpha\delta(x-x_0)$. Let the wave amplitudes immediately to the left and right of the delta be denoted by:
\begin{align}
\psi_L(x) &= A e^{ikx} + B e^{-ikx}, \quad x < x_0, \\
\psi_R(x) &= C e^{ikx} + D e^{-ikx}, \quad x > x_0.
\end{align}

The delta potential imposes two conditions at $x = x_0$: continuity of the wavefunction and a jump in its derivative proportional to the strength $\alpha$:
\begin{align}
\psi_R(x_0) &= \psi_L(x_0), \label{eq:cont_main} \\
\psi'_R(x_0) - \psi'_L(x_0) &= \frac{2m\alpha}{\hbar^2} \psi(x_0). \label{eq:jump_main}
\end{align}

These conditions relate the amplitudes on the two sides via a $2\times2$ transfer matrix $M$:
\begin{equation}
\begin{bmatrix}
C \\ D
\end{bmatrix}
= M 
\begin{bmatrix}
A \\ B
\end{bmatrix}.
\label{eq:transfer_relation}
\end{equation}

Solving \eqref{eq:cont_main} and \eqref{eq:jump_main} for $M$ yields:
\begin{equation}
M = \begin{bmatrix}
1 + \dfrac{i\alpha}{k} & \dfrac{i\alpha}{k}e^{-2ikx_0} \\[0.5em]
-\dfrac{i\alpha}{k}e^{2ikx_0} & 1 - \dfrac{i\alpha}{k}
\end{bmatrix}. \label{eq:M_single}
\end{equation}

The full derivation is well established which would take 4 pages here. See \cite{Griffiths2018} for reference. Throughout this work, we adopt natural units $\hbar = 1$ and $m = 1/2$ for computational simplicity. In these units, $k = \sqrt{E}$, and the dimensionless ratio $\alpha/k$ characterizes the relative strength of the delta potential compared to the particle's kinetic energy. Also, for scattering from the left, we set $A = 1$ (unit incident amplitude from the left) and $D = 0$ (no incoming wave from the right). Solving \eqref{eq:transfer_relation} gives the transmitted and reflected amplitudes:
\begin{equation}
C = \frac{1}{M_{11}}, \quad B = \frac{M_{21}}{M_{11}}.
\end{equation}
These correspond to the standard transmission and reflection coefficients $t = C$ and $r = B$, with corresponding probabilities:
\begin{equation}
T = |C|^2 = \frac{1}{|M_{11}|^2}, \quad R = |B|^2 = \left|\frac{M_{21}}{M_{11}}\right|^2,
\end{equation}
and $T + R = 1$ ensuring probability conservation.

For a system of $N$ delta potentials at positions $x_1 < x_2 < \dots < x_N$, the total transfer matrix is the ordered product \cite{Loran2015}:
\begin{equation}
M_{\text{total}} = M_N \cdots M_2 M_1, \label{eq:M_total}
\end{equation}
where each $M_i$ has the form \eqref{eq:M_single} with its own $\alpha_i$ and $x_i$. The overall transmission probability is:
\begin{equation}
T = \frac{1}{|(M_{\text{total}})_{11}|^2}. \label{eq:T_total}
\end{equation}
This compact formulation allows efficient computation of scattering spectra for arbitrary delta configurations, forming the basis of our numerical optimization.

\subsection{2-$\delta$ and 3-$\delta$ Systems}

The 2-$\delta$ system consists of two delta potentials at variable positions with strengths 
$\alpha_1$ and $\alpha_2$:
\begin{equation}
V_{2\delta}(x) = \alpha_1 \delta(x - x_a) + \alpha_2 \delta(x - x_b)
\end{equation}
with $x_a < x_b$. This gives the two-delta system four adjustable parameters: two strengths 
$(\alpha_1, \alpha_2)$ and two positions $(x_a, x_b)$, though the absolute position is irrelevant 
for transmission spectra---only the separation $\Delta x = x_b - x_a$ matters.

The transmission probability is obtained via the ordered product $M_{\text{total}} = M_2 M_1$, 
where each $M_i$ is given by Eq.~\eqref{eq:M_single}. The (1,1) element $A_{11}$ of the resulting matrix 
for the 2-$\delta$ system yields:
\begin{equation}
T_{2\delta}(k; \alpha_1, \alpha_2, \Delta x) =\frac{1}{|A_{11}|^2}= \frac{1}{\left[1 + \dfrac{\alpha_1\alpha_2}{k^2}(\cos(2k\Delta x) - 1)\right]^2 + \left[\dfrac{\alpha_1 + \alpha_2}{k} - \dfrac{\alpha_1\alpha_2}{k^2}\sin(2k\Delta x)\right]^2}
\label{eq:T2delta_main}
\end{equation}

The 3-$\delta$ system extends the scattering complexity with an additional barrier:
\begin{equation}
V_{3\delta}(x) = \beta_1 \delta(x - x_1) + \beta_2 \delta(x - x_2) + \beta_3 \delta(x - x_3)
\end{equation}
with positions constrained by $x_1 < x_2 < x_3$. This system has six adjustable parameters: three 
strengths $(\beta_1, \beta_2, \beta_3)$ and three positions $(x_1, x_2, x_3)$, though again only 
relative separations $\Delta x_{ij} = x_j - x_i$ matter for transmission. We denote these separations as 
$\Delta x_{12} = x_2 - x_1$, $\Delta x_{23} = x_3 - x_2$, and $\Delta x_{13} = x_3 - x_1$.

The transmission probability follows from the ordered product $M_{\text{total}} = M_3 M_2 M_1$ 
and takes the form:
\begin{equation}
T_{3\delta}(k; \beta_1, \beta_2, \beta_3, \Delta x_{12}, \Delta x_{23}) = \frac{1}{|B_{11}|^2} =\frac{1}{[\operatorname{Re}(B_{11})]^2 + [\operatorname{Im}(B_{11})]^2}
\label{eq:T3delta_main}
\end{equation}
where $B_{11}$ is the (1,1) element of the total transfer matrix, which can be expressed as 
$B_{11} = \operatorname{Re}(B_{11}) + i\operatorname{Im}(B_{11})$ with:
\begin{align*}
\operatorname{Re}(B_{11}) = &\ 1 - \frac{\beta_1\beta_2 + \beta_1\beta_3 + \beta_2\beta_3}{k^2} \\
& + \frac{\beta_1\beta_2}{k^2}\cos(2k\Delta x_{12}) + \frac{\beta_2\beta_3}{k^2}\cos(2k\Delta x_{23}) + \frac{\beta_1\beta_3}{k^2}\cos(2k\Delta x_{13}) \\
& - \frac{\beta_1\beta_2\beta_3}{k^3}\left[\sin(2k\Delta x_{12}) + \sin(2k\Delta x_{23}) - \sin(2k\Delta x_{13})\right]
\end{align*}
\begin{align*}
\operatorname{Im}(B_{11}) = &\ \frac{\beta_1 + \beta_2 + \beta_3}{k} - \frac{\beta_1\beta_2\beta_3}{k^3} \\
& - \frac{\beta_1\beta_2}{k^2}\sin(2k\Delta x_{12}) - \frac{\beta_2\beta_3}{k^2}\sin(2k\Delta x_{23}) + \frac{\beta_1\beta_3}{k^2}\sin(2k\Delta x_{13}) \\
& + \frac{\beta_1\beta_2\beta_3}{k^3}\left[\cos(2k\Delta x_{12}) + \cos(2k\Delta x_{23}) - \cos(2k\Delta x_{13})\right]
\end{align*}
Both transmission probabilities approach $1$ as $k$ grows larger, as expected.

\subsection{Necessary conditions for Exact Isospectrality}
\label{sec:exact_impossibility}

\begin{proposition}[Necessary conditions for exact isospectrality]
\label{prop:exact_impossible}
If $T_{2\delta}(k) = T_{3\delta}(k)$ for all large $k > 0$, then the following 
conditions must be satisfied:
\begin{enumerate}
\item The squared strength sums are equal:
\begin{equation}
\label{eq:strength_sum_app}
\alpha_1^2 + \alpha_2^2 = \beta_1^2 + \beta_2^2 + \beta_3^2.
\end{equation}

\item All pairwise products vanish:
\begin{equation}
\label{eq:pairwise_products_app}
\alpha_1\alpha_2 = \beta_1\beta_2 = \beta_2\beta_3 = \beta_1\beta_3 = 0.
\end{equation}
\end{enumerate}
\end{proposition}

The proof is given in Appendix~\ref{proof:proposition31}.

\begin{corollary}[Triviality of exact matches]
\label{cor:trivial_app}
Under the conditions of Proposition~\ref{prop:exact_impossible}, any exactly isospectral 
pair must be trivial in the following sense:
\begin{itemize}
\item The two-delta system has at most one non-zero delta ($\alpha_1\alpha_2 = 0$ 
implies at least one of $\alpha_1$ or $\alpha_2$ is zero).
\item The three-delta system has at most one non-zero delta (all pairwise products 
vanish, so at most one of $\beta_1,\beta_2,\beta_3$ can be non-zero).
\end{itemize}
Such systems effectively reduce to single-delta potentials.
\end{corollary}

\begin{figure}[htbp]
    \centering
    \includegraphics[width=1\textwidth]{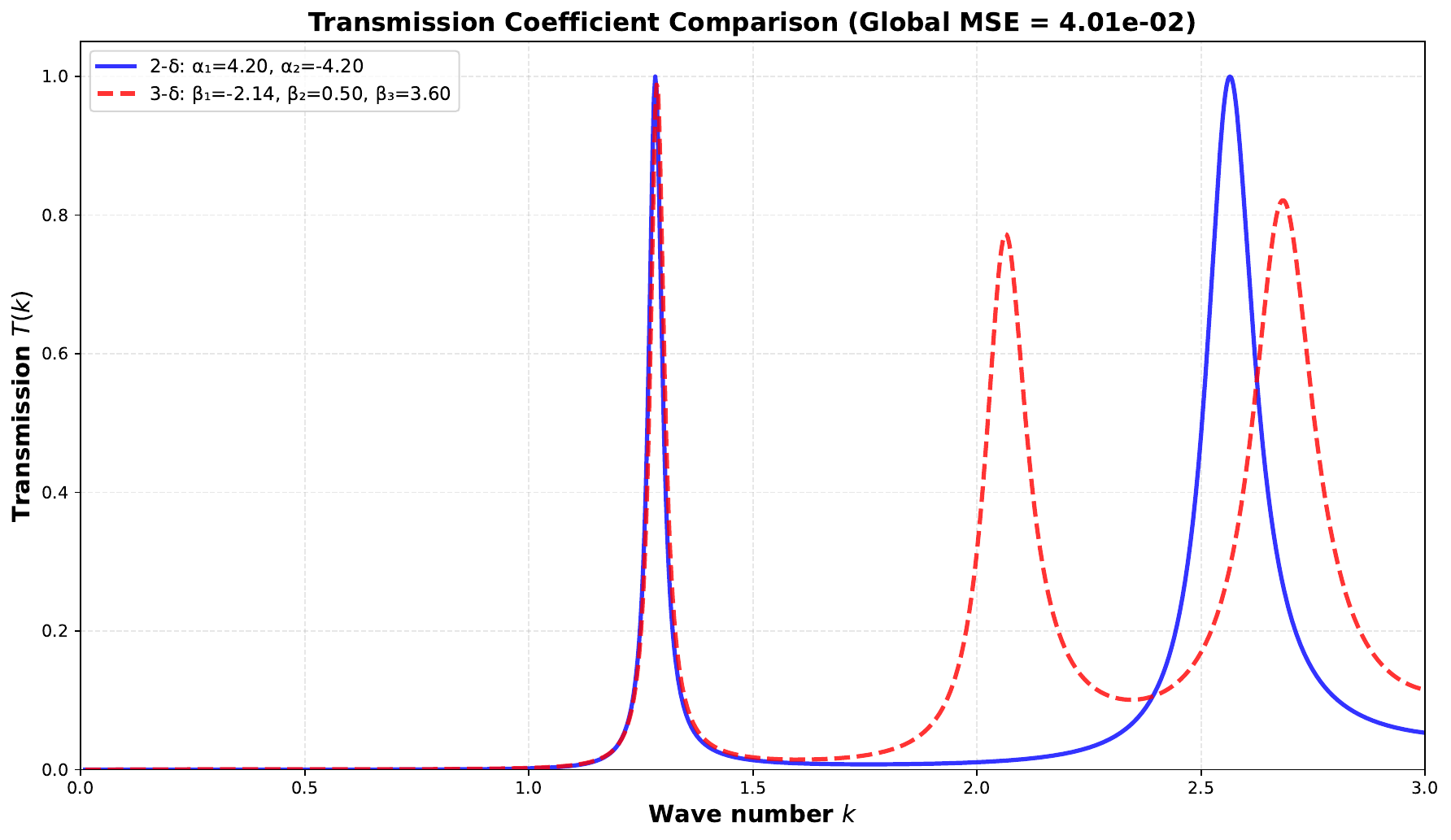}
    \caption{Failed isospectrality attempt for global optimization over $k \in [0.01, 3.0]$ achieves only MSE = $4.01 \times 10^{-2}$ (Positions are not shown explicitly for the emphasis on transmission). This validates Proposition 2.1 and motivates the windowed optimization strategy (Section 3), which achieves MSE $\sim 10^{-10}$ locally with the same positive-only constraints.}
    \label{fig:failed_matching}
\end{figure}

\subsection{Resonance Conditions for 2-$\delta$ Systems}

Perfect transmission ($T(k) = 1$) in the 2-$\delta$ system occurs when $\alpha_2 = -\alpha_1$, producing periodic resonances at
\begin{equation}
k_n = \frac{\pi n}{\Delta x}, \quad n = 1, 2, 3, \ldots
\label{eq:periodic_resonance}
\end{equation}
where $\Delta x = x_b - x_a$ is the separation between the delta potentials (derivation in Appendix~\ref{app:resonance}). This configuration generates an infinite series of evenly-spaced transmission peaks. Other resonance conditions exist (notably, equal-strength configurations $\alpha_1 = \alpha_2$ produce a single resonance at $k = \alpha_1$), but the periodic case is the only one that provides multiple resonances within finite energy ranges, making it the relevant configuration for windowed optimization (Section 3).

\section{Computation}

\subsection{Methodology}
To simultaneously match multiple resonances while enforcing physical constraints (bounded positive strengths, bounded positions), we employ windowed-optimization. We have $\alpha_2 = -\alpha_1$ and the transmission spectrum exhibits perfect transmission peaks at wave numbers:
\begin{equation}
k_n = \frac{\pi n}{\Delta x}, \quad n = 1, 2, 3, \ldots
\end{equation}
where $\Delta x = x_b - x_a$.

For each resonance peak $k_n$ within the range $k \in [0.01, 3.0]$, we construct a symmetric optimization window $W_n = [k_n - w, k_n + w]$, where the half-width $w$ is determined by:
\begin{equation}
w = \begin{cases}
\frac{1}{2}\langle k_{n+1} - k_n \rangle & \text{if multiple resonances exist} \\
\frac{k_n}{2} & \text{if only one resonance}
\end{cases}
\end{equation}
where $\langle k_{n+1} - k_n \rangle$ is the average spacing between consecutive resonance peaks. This ensures all windows have comparable width and each resonance peak is centered within its respective window.

For each window, we optimize the parameters of a 3-$\delta$ system to minimize the mean squared error (MSE) between its transmission spectrum $T_{3\delta}(k)$ and that of the target 2-$\delta$ system $T_{2\delta}(k)$:
\begin{equation}
\text{MSE}_n = \frac{1}{M} \sum_{j=1}^{M} \left[ T_{2\delta}(k_j) - T_{3\delta}(k_j) \right]^2
\end{equation}
where $\{k_1, k_2, \ldots, k_M\}$ are $M = 200$ uniformly spaced points within window $W_n$.

All numerical computations were performed using Python with the NumPy \cite{Harris2020} and SciPy \cite{Virtanen2020} libraries. The optimization employed the differential evolution algorithm from SciPy's \texttt{scipy.optimize} module. Results were visualized using Matplotlib \cite{Hunter2007}.

\subsection{Parametrization and Constraints}

The transmission spectrum $T(k)$ depends only on the relative positions of the delta functions, not their absolute locations. We exploit this by fixing the first position $x_1 = x_a$ (matching the leftmost delta of the target 2-$\delta$ system), reducing the parameter space from six to five independent variables while preserving all physically distinct configurations.

For a 3-$\delta$ system, we optimize five parameters: three potential strengths $\{\beta_1, \beta_2, \beta_3\}$ and two position spacings $\{\Delta x_{12}, \Delta x_{23}\}$, where $\Delta x_{12} = x_2 - x_1$ and $\Delta x_{23} = x_3 - x_2$. Parameters are ordered as $(\beta_1, \beta_2, \beta_3, \Delta x_{12}, \Delta x_{23})$ in the optimization vector. The first position $x_1$ is fixed to the value $x_a$ of the target 2-$\delta$ system to eliminate translational invariance, and the remaining positions are determined by:
\begin{itemize}
    \item $x_1 = x_a$ (fixed)
    \item $x_2 = x_1 + \Delta x_{12}$, where $\Delta x_{12} \in [d_{\text{min}}, 5]$
    \item $x_3 = x_2 + \Delta x_{23}$, where $\Delta x_{23} \in [d_{\text{min}}, 5]$
\end{itemize}
with minimum separation $d_{\text{min}} = 0.3$ to ensure physical significance and prevent numerical instabilities.

All potential strengths are constrained to $\beta_i \in [0.5, 2|\alpha_1|]$. The reason for the lower bound is physical significance. And the reason for the upper bound is in section \ref{sec: strength constraint}.

\subsection{Optimization Algorithm}

We employ the differential evolution algorithm with the following parameters:
\begin{itemize}
    \item \textbf{Population size}: 20 individuals
    \item \textbf{Maximum iterations}: 800
    \item \textbf{Convergence tolerances}: $\text{atol} = \text{tol} = 10^{-10}$
    \item \textbf{Random seed}: 42 (for reproducibility)
\end{itemize}

\subsection{Configuration Space}

The number of resonances (and thus optimization windows) depends only on the separation of the target 2-$\delta$ system. For the configurations studied in this work with $\alpha_1 = -\alpha_2 \in \{2, 3, 4\}$ and varying separations $\Delta x$, we focus on configurations with 1 to 5 resonances within $k \in [0.01, 3.0]$. Each resonance produces an independently optimized 3-$\delta$ configuration, resulting in a piecewise approximation to the 2-$\delta$ transmission spectrum with automatic adaptation to the target system's spectral structure.

\section{Results and Analysis}

We systematically studied windowed approximations across coupling strengths ($\alpha = \pm 2, \pm 3, \pm 4$) and resonance configurations (1-5 resonances). Under the strength constraint ($0.5<\beta_i \leq 2|\alpha_1|$) for 3-$\delta$ systems, we observe consistent resonance matching across all energy ranges.

\begin{figure}[htbp]
    \centering
    \includegraphics[width=\textwidth]{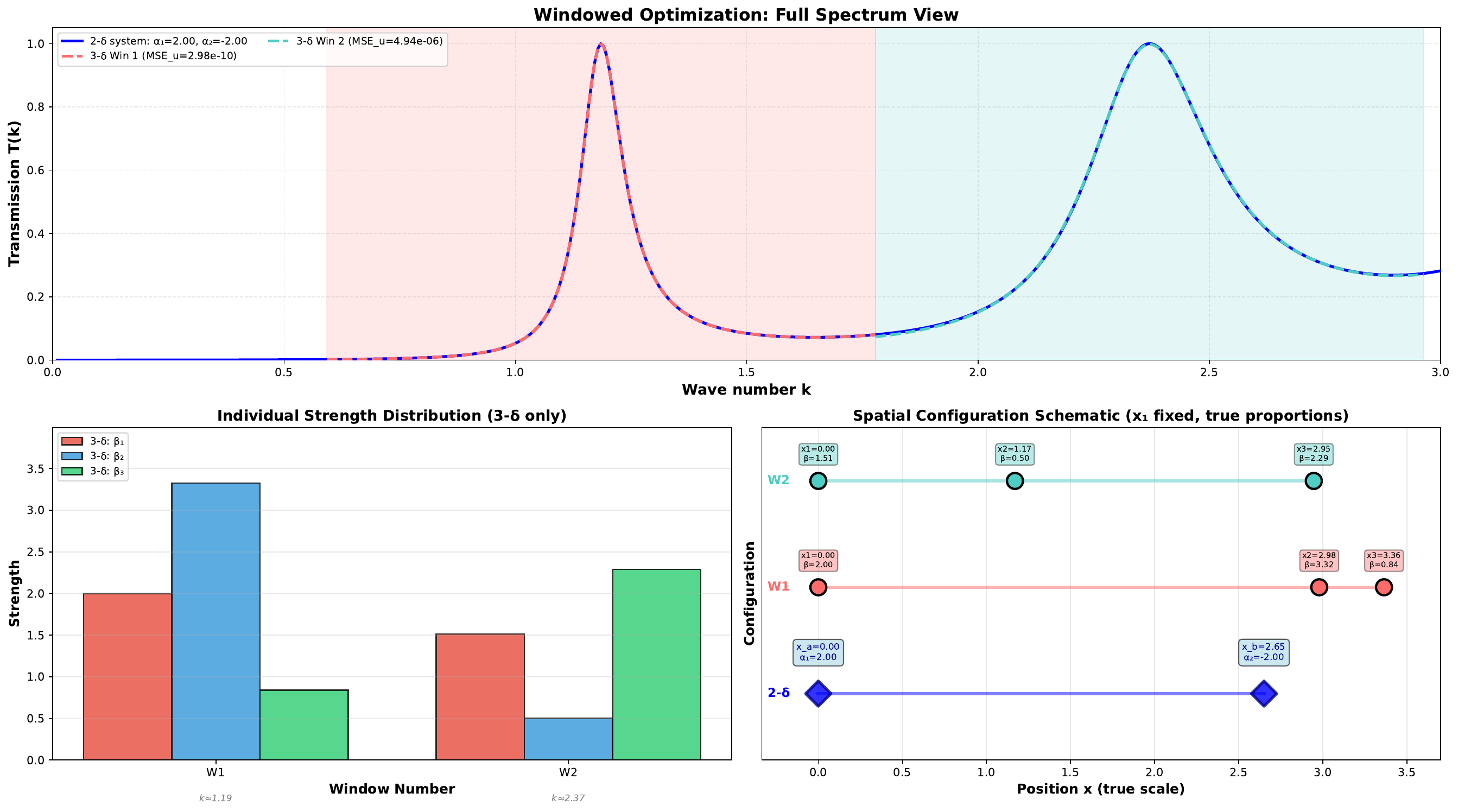}
    \caption{Two-resonance configuration with $\alpha = \pm 2$, $\Delta x = 2.65$. Resonances at $k_1 \approx 1.19$ and $k_2 \approx 2.37$ with MSE: $2.98 \times 10^{-10}$ (W1) and $4.94 \times 10^{-6}$ (W2). The strength distribution shows all $\beta_i$ values remain within the $2|\alpha_1| = 4$ bound.}
    \label{fig:2peaks}
\end{figure}

\subsection{Representative Cases: 2 and 5 Resonances}

Consider first a two-resonance configuration with coupling strength $\alpha = \pm 2$ and delta separation $\Delta x = 2.65$, producing resonances at $k_1 \approx 1.19$ and $k_2 \approx 2.37$ (Fig.~\ref{fig:2peaks}). Both windows achieve excellent MSE values of $2.98 \times 10^{-10}$ and $4.94 \times 10^{-6}$ respectively, with the 3-$\delta$ approximations accurately reproducing peak heights, widths, and baseline transmission throughout each energy range.

\begin{figure}[htbp]
    \centering
    \includegraphics[width=\textwidth]{5_peaks.pdf}
    \caption{Five-resonance configuration with $\alpha = \pm 3$, $\Delta x = 5.97$. Resonances at $k_1 \approx 0.53$, $k_2 \approx 1.05$, $k_3 \approx 1.58$, $k_4 \approx 2.10$, $k_5 \approx 2.63$ with MSE values ranging from $1.61 \times 10^{-8}$ to $5.22 \times 10^{-5}$. All strengths satisfy $\beta_i \leq 2|\alpha_1| = 6$.}
    \label{fig:5peaks}
\end{figure}

The optimizer adopts distinct spatial strategies for each window. The lower-energy window (W1) employs a balanced distribution with strengths $\beta_1 = 2.00$, $\beta_2 = 3.32$, $\beta_3 = 0.84$ spanning $\Delta x_{13} = 3.36$. The higher-energy window (W2) shows $\beta_1 = 1.51$, $\beta_2 = 0.50$, $\beta_3 = 2.29$ with spatial extent $\Delta x_{13} = 2.95$. Notably, all individual strengths remain below the $2|\alpha_1| = 4$ threshold, with the maximum value $\beta_2^{(W1)} = 3.32$ approaching but not exceeding this bound.

To demonstrate scalability, we examine a five-resonance configuration with $\alpha = \pm 3$ and $\Delta x = 5.97$, producing resonances at $k_1 \approx 0.53$, $k_2 \approx 1.05$, $k_3 \approx 1.58$, $k_4 \approx 2.10$, and $k_5 \approx 2.63$ (Fig.~\ref{fig:5peaks}). The five optimization windows achieve MSE values of $1.61 \times 10^{-8}$ (W1), $9.37 \times 10^{-7}$ (W2), $2.80 \times 10^{-7}$ (W3), $5.11 \times 10^{-8}$ (W4), and $5.22 \times 10^{-5}$ (W5), demonstrating consistent approximation quality across the extended energy range.

The optimizer employs window-specific spatial strategies. Window 1 (lowest energy) uses strengths $\beta_1 = 5.99$, $\beta_2 = 0.50$, $\beta_3 = 3.05$ spanning $\Delta x_{13} = 5.63$. Window 2 adopts $\beta_1 = 0.50$, $\beta_2 = 2.60$, $\beta_3 = 4.30$ over $\Delta x_{13} = 5.42$. Window 3 shows $\beta_1 = 0.50$, $\beta_2 = 2.58$, $\beta_3 = 3.63$ with $\Delta x_{13} = 7.80$. Window 4 employs $\beta_1 = 3.39$, $\beta_2 = 2.23$, $\beta_3 = 0.71$ spanning $\Delta x_{13} = 7.48$. Window 5 uses $\beta_1 = 6.00$, $\beta_2 = 2.21$, $\beta_3 = 2.07$ over $\Delta x_{13} = 4.77$. Critically, all strengths remain below the $2|\alpha_1| = 6$ bound, with the maximum value $\beta_1^{(W1)} = 5.99$ approaching but not exceeding this threshold while maintaining excellent spectral matching across all five windows.

While we have our lower bound of $\beta_i$ to be 0.5 and one of the $\beta_i$ naturally tends towards the lower bound to converge into the original 2-$\delta$ system, we noticed that the first window in figure \ref{fig:2peaks} and the fourth and fifth windows in figure \ref{fig:5peaks} do not have the minimum potential strengths of 0.5. This mean for some resonance of 2-$\delta$ system, it is possible to approximate with 3-$\delta$ system that is entirely different.

\subsection{Performance Trends Across System Complexity}

\begin{figure}[htbp]
    \centering
    \includegraphics[width=\textwidth]{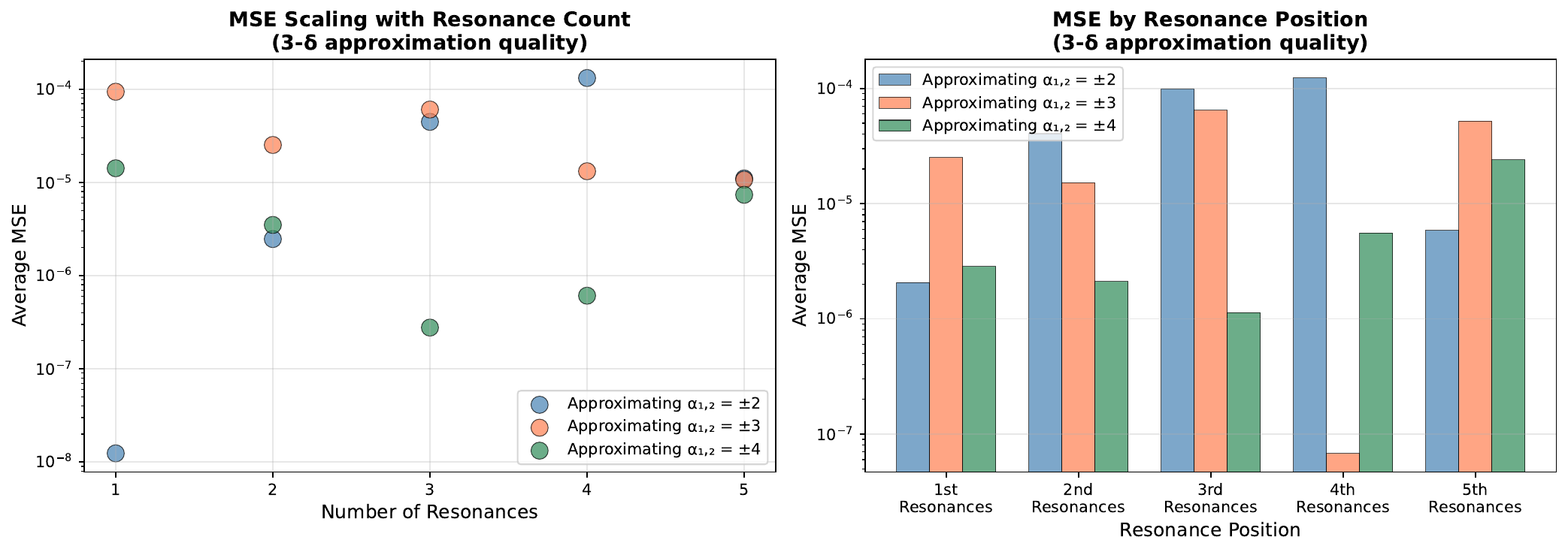}
    \caption{Left: Average MSE versus number of resonances for different coupling strengths. Right: Average MSE organized by resonance position across all systems. Under the $2|\alpha_1|$ strength bound, performance remains consistently strong across all resonance counts and positions. Here, `Average MSE' denotes the mean MSE computed over all windows within each 2-delta system.}
    \label{fig:mse_analysis}
\end{figure}

Analysis of MSE versus system complexity (Fig.~\ref{fig:mse_analysis}) shows no clear systematic pattern, but most configurations achieve average MSE below $10^{-4}$, which is sufficient for practical purposes. The right panel demonstrates that the $2|\alpha_1|$ strength bound is proven to be both necessary and sufficient for uniform performance across all resonance positions, with typical values ranging from $10^{-6}$ to $10^{-4}$.
 The method performs consistently well for systems with 1-5 resonances spanning $k \in [0.01, 3.0]$. 

\subsection{Strength constraints for successful approximation}
\label{sec: strength constraint}
Through numerical experimentation, we found that constraining the optimized coupling strengths to $0.5 \leq  |\beta_i| \leq 2|\alpha_1|$ was necessary for successful approximations (i.e., MSE $< 10^{-3}$). Tighter bounds such as $0.5 \leq|\beta_i| \leq |\alpha_1|$ succeeded for higher-order resonances but systematically failed to match first resonances at low $k$. 

This factor of 2 appears to be a practical requirement of the optimization procedure rather than a deep physical constraint. It may simply provide sufficient parameter space for the optimizer to find adequate solutions across all energy ranges. The relationship to the original coupling strength $|\alpha_1|$ provides a convenient scaling rule, but the precise value likely depends on the optimization method and energy window rather than fundamental physics.

\section{Discussion and Limitation}
\label{sec:discussion}

The approximation quality depends critically on optimization freedom: overly restrictive constraints $\beta_i \leq |\alpha_1|$ prevent faithful low-energy matching, while moderate relaxation $\beta_i \leq 2|\alpha_1|$ enables excellent performance uniformly.

Several questions merit future investigation. How does the required strength bound scale when approximating $N$-delta systems with $(N+1)$-delta configurations? Can analytical arguments predict optimal constraints from fundamental scattering properties? Furthermore, although our optimization identified effective spatial configurations 
\((\Delta x_{12}, \Delta x_{23})\) for each window, we did not systematically analyze 
how these arrangements correlate with resonance energy or target system parameters. 
Future work could explore whether optimal spatial patterns emerge.

Proposition~\ref{prop:exact_impossible} (proved in Appendix~\ref{proof:proposition31}) establishes that exact global isospectrality is impossible for non-trivial configurations, and we conjecture it generalizes: for finite delta collections with $m \neq n$ deltas, exact isospectrality is impossible unless at most one delta in each configuration has nonzero strength. 

Lastly, the method produces different 3-$\delta$ configurations for each energy window rather than a single configuration that approximates the entire spectrum. Practical implementations requiring a fixed potential across all energies would need alternative approaches.

\section{Conclusions}
\label{sec:conclusions}

We developed a windowed optimization method to approximate 2-$\delta$ potentials with opposite sign coupling strengths using positive-only 3-$\delta$ potentials with appropriate constraints. Proposition \ref{prop:exact_impossible} established necessary conditions for exact isospectrality which implies impossibility of exact spectrum matching across all $k$. Nevertheless, numerical experiments identified the minimal constraint set for practical approximation. The approach succeeds uniformly across system complexities (1-5 resonances) and coupling strengths ($|\alpha| = 2, 3, 4$), achieving MSE values from $10^{-8}$ to $10^{-4}$.

The strength bound $\beta_i \leq 2|\alpha_1|$ proves both necessary and sufficient across all tested configurations, with coupling-strength independence suggesting robust scalability. These results establish both the practical utility and fundamental constraints of approximate isospectrality in quantum transmission problems, with implications for inverse scattering theory and potential engineering applications.

\section*{Acknowledgments}
The Python code used for optimization and figure generation in this work is available at: 
\url{https://github.com/nawsailabang/Resonance-Matching-for-2-delta-and-3-delta-in-1D-Quantum-Scattering}

\appendix

\subsection{Proof of Proposition~\ref{prop:exact_impossible}}
\label{proof:proposition31}
\begin{proof}
For $T_{2\delta}(k) = T_{3\delta}(k)$ to hold for all sufficiently large $k$, the 
expansions in powers of $u = 1/k$ must match term by term. From equations \eqref{eq:T2delta_main} and \eqref{eq:T3delta_main}, we have:

\[
|A_{11}|^2 = 1 + \left[(\alpha_1+\alpha_2)^2 + 2\alpha_1\alpha_2(\cos(2k\Delta x)-1)\right]u^2 + O(u^3)
\]
\begin{align*}
|B_{11}|^2 = 1 &+ \bigg[(\beta_1+\beta_2+\beta_3)^2 - 2(\beta_1\beta_2 + \beta_1\beta_3 + \beta_2\beta_3) \\
&+ 2\big(\beta_1\beta_2\cos(2k\Delta x_{12}) + \beta_2\beta_3\cos(2k\Delta x_{23}) + \beta_1\beta_3\cos(2k\Delta x_{13})\big)\bigg]u^2 + O(u^3)
\end{align*}

\textbf{Matching $u^0$ terms:} Both expansions begin with $1$, trivially satisfied.

\textbf{Matching $u^2$ terms:} Equating the coefficients of $u^2$ from equations \eqref{eq:T2delta_main} and \eqref{eq:T3delta_main}, we require:
\begin{align*}
&(\alpha_1+\alpha_2)^2 + 2\alpha_1\alpha_2(\cos(2k\Delta x)-1) \\
&= (\beta_1+\beta_2+\beta_3)^2 - 2(\beta_1\beta_2 + \beta_1\beta_3 + \beta_2\beta_3) \\
&\quad + 2\big(\beta_1\beta_2\cos(2k\Delta x_{12}) + \beta_2\beta_3\cos(2k\Delta x_{23}) + \beta_1\beta_3\cos(2k\Delta x_{13})\big).
\end{align*}

For generic values of $\Delta x$, $\Delta x_{12}$, $\Delta x_{23}$, $\Delta x_{13}$ with no special rational relationships, the trigonometric terms $\cos(2k\Delta x)$, $\cos(2k\Delta x_{12})$, $\cos(2k\Delta x_{23})$, and $\cos(2k\Delta x_{13})$ oscillate independently as $k \to \infty$. Since the equality must hold for all sufficiently large $k$, the coefficients of each independent oscillatory term must vanish separately:
\begin{equation}
\alpha_1\alpha_2 = 0, \quad \beta_1\beta_2 = \beta_2\beta_3 = \beta_1\beta_3 = 0.
\label{eq:pairwise_products_app}
\end{equation}

\begin{remark}
One might ask whether special geometric relationships (e.g., $\Delta x_{12} = \Delta x$, $\Delta x_{23} = \Delta x/2$) could allow trigonometric terms to cancel while keeping nonzero coefficients. However, such configurations would require satisfying infinitely many constraints (one for each value of $k$) using only finitely many free parameters ($\Delta x_{12}$, $\Delta x_{23}$, $\Delta x_{13}$), which is generically impossible. Thus, condition \eqref{eq:pairwise_products_app} holds for all but a measure-zero set of configurations.
\end{remark}

With the trigonometric terms eliminated, the remaining constant terms give:
\[
(\alpha_1+\alpha_2)^2 - 2\alpha_1\alpha_2 = (\beta_1+\beta_2+\beta_3)^2 - 2(\beta_1\beta_2 + \beta_1\beta_3 + \beta_2\beta_3).
\]
Using the identity $(\sum a_i)^2 - 2\sum_{i<j} a_i a_j = \sum a_i^2$, this simplifies to 
condition \eqref{eq:strength_sum_app}:
\begin{equation}
\alpha_1^2 + \alpha_2^2 = \beta_1^2 + \beta_2^2 + \beta_3^2.
\label{eq:strength_sum_app}
\end{equation}
\end{proof}

\subsection{Derivation of Resonance Conditions for 2-$\delta$ system}
\label{app:resonance}

Let $\Delta x = x_b - x_a$ denote the separation between the two delta functions. Perfect resonance with $T = 1$ requires that the denominator of Eq.~\eqref{eq:T2delta_main} equals unity:
\begin{equation}
\left[1 + \frac{\alpha_1\alpha_2}{k^2}(\cos(2k\Delta x) - 1)\right]^2 + \left[\frac{\alpha_1 + \alpha_2}{k} - \frac{\alpha_1\alpha_2}{k^2}\sin(2k\Delta x)\right]^2 = 1
\end{equation}

We denote the two terms as:
\begin{align}
A &= 1 + \frac{\alpha_1\alpha_2}{k^2}(\cos(2k\Delta x) - 1)\\
B &= \frac{\alpha_1 + \alpha_2}{k} - \frac{\alpha_1\alpha_2}{k^2}\sin(2k\Delta x)
\end{align}

The condition $A^2 + B^2 = 1$ defines a circle in the $(A,B)$ plane with four candidate solutions: $(A,B) = (\pm1, 0)$ and $(A,B) = (0, \pm1)$. However, only Case 1 yields the periodic resonance structure used in this work.

\textbf{Case 1:} $A = 1, B = 0$

From $A = 1$, we obtain $\cos(2k\Delta x) = 1$, giving:
\begin{equation}
k_n = \frac{\pi n}{\Delta x}, \quad n = 1, 2, 3, \ldots
\end{equation}

From $B = 0$ with $\sin(2k\Delta x) = 0$, we require:
\begin{equation}
\frac{\alpha_1 + \alpha_2}{k} = 0 \quad \Rightarrow \quad \alpha_2 = -\alpha_1
\end{equation}

\textit{Result:} Periodic resonances at $k_n = \pi n/\Delta x$ when the delta functions have equal and opposite strengths. This produces an infinite series of transmission peaks and is the physically relevant case for windowed optimization.

\textbf{Cases 2, 3, and 4:} The remaining cases yield only limited resonance structures: $(A,B) = (-1, 0)$ gives $k = 0$ or negative $k$; $(A,B) = (0, 1)$ requires $\alpha_1 = \alpha_2$ and produces a single isolated resonance at $k = \alpha_1$; and $(A,B) = (0, -1)$ yields only negative $k$ values. None of these cases provide the multiple resonances needed for the optimization approach in Section 3.

\section*{Acknowledgments}
I would like to thank Dr. Sicong Zhang, assistant professor of applied mathematics at Beijing Normal-Hong Kong Baptist University, for the supervision of this project.

\bibliographystyle{plain}  
\bibliography{references}

\end{document}